\documentclass[aps]{revtex4}
\usepackage{graphicx}
\usepackage{epic}
\usepackage{eepic}
\newcommand{\be}{\begin{equation}}
\newcommand{\ee}{\end{equation}}
\newcommand{\bea}{\begin{eqnarray}}
\newcommand{\eea}{\end{eqnarray}}

\newcommand{\hs}[1]{\hspace{#1 mm}}

\begin{document}

\baselineskip=.50cm

\title{A Note on Supergravity Solutions for Partially Localized \\ 
Intersecting Branes}
\author{S. Arapoglu}
\email[e-mail:]{arapoglu@boun.edu.tr}
\affiliation{Bogazici University, Department of Physics\\ 
34342, Bebek, Istanbul, Turkey}
\author{N.S. Deger}
\email[e-mail:]{deger@gursey.gov.tr}
\author{A. Kaya}
\email[e-mail:]{kaya@gursey.gov.tr}
\affiliation{Feza Gursey Institute,\\
Cengelkoy, 81220, Istanbul, Turkey}
\date{\today}
\begin{abstract}
\baselineskip=.40cm 
Using the method developed by Cherkis and Hashimoto we construct
partially localized  $D3\perp D5(2)$, $D4\perp D4(2)$ and $M5\perp M5(3)$
supergravity solutions  where one of the harmonic functions is given
in an integral form. This is a generalization of the already known
near-horizon solutions. The method fails for certain intersections
such as $D1 \perp D5 (1)$ which is consistent with the previous no-go
theorems. We point out some possible ways of bypassing these results.   
\end{abstract}
\maketitle

\section{Introduction}

There has been considerable interest in constructing intersecting brane
solutions in the past (see \cite{rev1, rev2, rev3} for review). The problem is
completely solvable if one assumes that the solution depends only on
overall transverse directions. However relaxing this condition complicates
it considerably. If the metric is chosen to be in some specific form 
(which is inspired by harmonic function rule \cite{har1, har2, har3}) 
then it is easy to see that one of the brane has to be delocalized
\cite{pope}, i.e., its harmonic function is independent of the
directions along the other brane's worldvolume. This is not a
restriction if the smaller brane is contained in the bigger one;
otherwise these type of solutions are said to be {\it partially
localized} (see figure 1). Explicit intersections have been found by further
restricting to the near-horizon of the delocalized brane
\cite{tseytlin, has2, youm, loewy}. Recently Cherkis and Hashimoto
\cite{hashimoto} were able to remove this restriction for $D2 \perp D6
(2)$ intersection which allowed them to analyze the system in the
near-horizon region of $D2$ instead of $D6$ which has some important
applications in $AdS/CFT$ duality. This method has been further applied to
construct $D1 \perp NS5 (0)$ intersection in \cite{nas1} and 
$D4 \perp D8 (4)$ intersection in \cite{nas2}. 

The approach of \cite{hashimoto}, which we adopt in this paper,  
is similar to the the technique used in \cite{mar1, mar2, mar3} to
prove no-hair theorems for $p$-branes. 
It is a generic feature of intersecting brane
configurations that the differential equations involving the
metric functions are linear and separable. This lets one to
apply Fourier transformation techniques which allows the construction
of the harmonic function as an integral expression. This can be
evaluated numerically if desired and it is a generalization of the 
near-horizon solutions given in \cite{youm}. (See also \cite{fay1,
fay2, fay3}.)   

As we will discuss below, this method fails when the overall
transverse space, $(n+2)$, is four or higher dimensional. When $n>2$,
for instance as in  the case of $D0\perp D4(0)$, the radial
dependence of the metric functions cannot be determined in terms of
elementary or known functions. On the other hand, for $n=2$
there is a generic spontaneous delocalization when the branes are
forced to be placed on top of each other. This is consistent with the 
previous no-go results for a full localization in such brane systems
\cite{mar1, mar2, mar3}.  

\begin{figure}
\centerline{
\includegraphics[width=14.0cm]{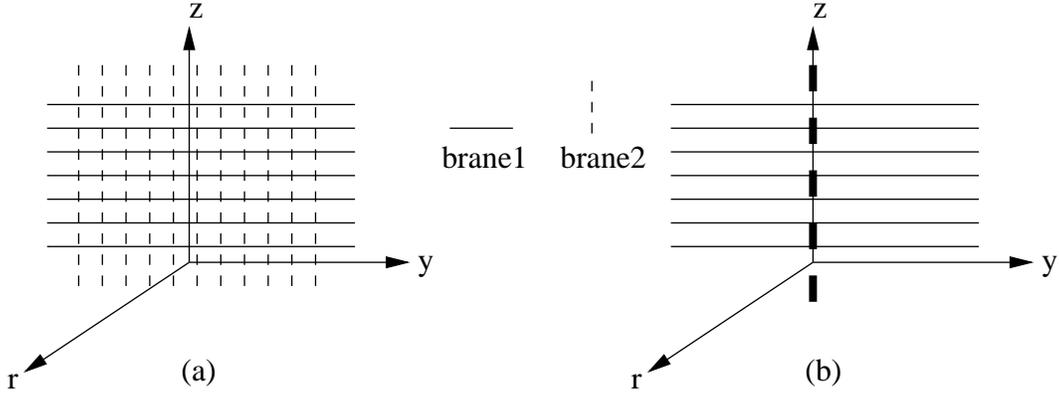}}
\caption{Delocalized (a) versus partially localized (b) brane
intersections. The first brane has the world-volume coordinates $(x,y)$
and the second one is oriented along $(x,z)$ directions ($x$ coordinate is
suppressed in the figure). In (a),  branes are smeared along $z$ and
$y$ coordinates, respectively. In (b), only the first brane is smeared
and the second brane is located at $y=0$. }
\end{figure}

\section{Solutions}

Let us start with an intersection of two D-branes which has the following 
metric
\be\label{met1}
ds^2= H_1^{-1/2}H_2^{-1/2}dx^\mu dx^\mu +H_1^{-1/2}H_2^{1/2}d\vec{y}.d\vec{y} +
H_1^{1/2}H_2^{-1/2}d\vec{z}.d\vec{z} + H_1^{1/2}H_2^{1/2}d\vec{r}.d\vec{r},
\ee
where $(x^\mu,\vec{y})$ and $(x^\mu,\vec{z})$ are the world-volume
coordinates of the first and the second branes which are characterized by the
``harmonic'' functions $H_1(\vec{z},\vec{r})$ and
$H_2(\vec{y},\vec{r})$. Changing the powers 
of metric functions the same metric can be thought to describe
intersection of two M-branes. We follow the usual brane terminology;
$x$ is a common brane coordinate, $\vec{y}$ and $\vec{z}$ 
are relative transverse directions and $\vec{r}$ coordinates
parameterize the overall transverse directions.  We assume that the
brane functions do not depend on the corresponding brane
coordinates. To have a localized solution one should have 
\bea
\lim_{|\vec{z}|\to\infty}H_1&\to& 1, \\    
\lim_{|\vec{y}|\to\infty}H_2&\to& 1.
\eea
The harmonic functions satisfy the following differential equations
\cite{pope} 
\bea
(\partial^2_{\vec{r}}+H_2\partial^2_{\vec{z}})H_1&=& 
q_1 \, \delta(\vec{r}) \, \delta(\vec{z}),\label{1}\\
(\partial^2_{\vec{r}}+H_1\partial^2_{\vec{y}})H_2&=&
q_2\, \delta(\vec{r}) \, \delta(\vec{y}),\label{2}\\
\partial_{\vec{z}}H_1\,\partial_{\vec{y}}H_2&=&0,\label{3}
\eea 
where the branes are assumed to be located at $\vec{r}=\vec{z}=0$ and
$\vec{r}=\vec{y}=0$, respectively.
The last equation indicates that either $\partial_{\vec{z}}H_1=0$ or
$\partial_{\vec{y}}H_2=0$, i.e. one of the branes should be delocalized along
the other brane directions. Without loss of generality we take it
to be the first brane.  Assuming spherical symmetry, (\ref{1}) gives
(up to an irrelevant numerical factor) 
\be\label{h1}
H_1=1+\frac{q_1}{r^n},
\ee
where $(n+2)$ is the dimension of the $\vec{r}$-space and $q_1$ is the brane
charge. For the special intersection where the second brane is
located inside the first one, $\vec{z}$ coordinates should be ignored. For
this case $H_1$ depends only on $\vec{r}$ and (\ref{3}) is satisfied
trivially. This corresponds to a full localization. When $H_1$ is
solved as in (\ref{h1}), the solutions of (\ref{2}) has been studied
in certain  limits. For instance, near horizon geometries where one
can take $H_1\sim r^{-n}$ were constructed in \cite{youm}. 
Following \cite{hashimoto}, to solve (\ref{2}) exactly, we use a Fourier
transformation in the $\vec{y}$ space to write
\bea
H_2&=&1+q_2\int d^m p\,\,e^{i\,\vec{p}.\vec{y}}\,\,H_p(r),\nonumber\\
&=&1+q_2\,\int_0^\infty \, dp\,\int_0^\pi d\theta\,(\sin\theta)^{m-2}\,p^{m-1}\,\Omega_{m-2}\,e^{ipy\cos\theta}\,\,H_p(r),\label{t}
\eea   
where $q_2$ is the brane charge, $m$ denotes dimension of the
$\vec{y}$ space, and $\Omega_{m-2}$ 
is the volume of the unit $(m-2)$-dimensional sphere with
$\Omega_0=1$. The above formula is valid when $m>1$ and for $m=1$ the
second step is unnecessary. For technical convenience, 
we first locate the second brane at $\vec{r}=\vec{r}_0$ and
then take $\vec{r}_0\to 0$ limit. Then, from (\ref{2}) and (\ref{t}) one finds
\be\label{rd}
\left[\frac{d^2}{dr^2}+\frac{n+1}{r}\frac{d}{dr}-
p^2(1+\frac{q_1}{r^n})\right]H_p(r)=\,q_2\,\frac{\delta(r-r_0)}{r^{n+1}}.
\ee
For each $m$, the $\theta$ integral in (\ref{t}) can be carried out
easily. Therefore, if one can solve (\ref{rd}) 
$H_2$ can be determined in an integral form which can be
evaluated numerically if wanted. Now let us discuss possible solutions
of (\ref{rd}):

\
\

\leftline{${\bf n\geq3:}$}

\
\

It turns out (\ref{rd}) cannot be
solved in terms of elementary functions (at least to our
knowledge). Recalling that $(n+2)$ is the dimension of the overall
transverse space, this corresponds to the intersections like $D0\perp
D4(0)$ or $M2\perp M2(0)$. 

\
\

\leftline{${\bf n=2:}$} 

\
\

The prototype of this case that we will consider is $D1\perp D5(1)$
intersection. However, since our arguments are based on the
$r$-dependence of the harmonic functions (which is fixed by $n$), our
conclusions apply intersections like $D2\perp D4(1)$ and $M2
\perp M5(1)$ as well.  
Even though, there are no-go theorems for the existence of
a localized solution \cite{mar1}-\cite{mar3}, for completeness we
will investigate this case too in order to emphasize the origin of the
difficulty. We will also propose some possible ways to resolve this. 
The solution to (\ref{rd}) which is both regular at $r=0$ and $r=\infty$ 
can be written as (we demand regularity at $r=0$ since we are mainly
interested in $r_0\to0$ limit)
\be\label{k}
H_p(r)=\begin{cases}{c_p(r_0)\,r^{-1}\,K_{\nu}(pr),\hs{5}r>r_0, \cr\cr
d_p(r_0)\,r^{-1}\,I_{\nu}(pr),\hs{6.5}r<r_0,}
\end{cases}
\ee
where $K_\nu$ and $I_\nu$ are the modified Bessel function with 
$\nu=\sqrt{1+q_1\,p^2}$ and $[c_p(r_0),d_p(r_0)]$ are constants. The
continuity at $r=r_0$ gives
\be\label{c1}
c_p(r_0)K_\nu(pr_0)=d_p(r_0)I_\nu(pr_0).
\ee
Using this in the condition imposed by the presence of the delta
function source at $r=r_0$ one obtains
\be
c_p(r_0)\,p\,W\{I_\nu(pr_0),K_\nu(pr_0)\}\,=\,q_2\,r_0^{-2}\,I_\nu(pr_0), 
\ee
where $W$ is the Wronskian with respect to the argument which is
equal to $-1/(pr_0)$. This implies
$c_p(r_0)=-q_2\,I_\nu(pr_0)/r_0$. In the $r_0\to0$ limit $c_p(r_0)\sim
r_0^{(\nu-1)}\to0$ which indicates spontaneous delocalization. This is
the essence of the trouble in $D1/D5$ localized solution. Physically,
as the separation goes to zero the $D1$-brane charge spreads over the
$D5$-brane.  

Now we would like to point out two possible ways of resolving this difficulty
although we could not establish a clear cut result. Firstly, there may
be a subtlety in taking $r_0\to0$ limit. Namely, a localized 
intersection when branes are coincident may not be continously reached
from a separated brane configuration. If so, then one should solve
(\ref{2}) directly without assuming any separation between the
branes. In this case, one  finds that $H_p(\vec{r})$ in (\ref{t})
obeys 
\be
\left[\partial_{\vec{r}}^{2}-p^2(1+\frac{q_1}{|\vec{r}|^2})\right]\,H_p(\vec{r})=\,q_2\,\delta(\vec{r}). \label{preint}
\ee 
Fourier expanding $H_p(\vec{r})$ as 
\be
H_p(\vec{r})=\int d^4 v \,e^{i\vec{r}.\vec{v}}\,h_p(\vec{v}),
\ee
(\ref{preint}) gives
\be\label{sint}
(2\pi)^{4}\,(|\vec{v}|^2+|\vec{p}|^2)\,h_p(\vec{v})\,+\,4\pi^2 p^2 q_1\,\int d^4 v' \frac{h_p(\vec{v'})}{|\vec{v}-\vec{v'}|^2}\,= \,-q_2.
\ee
Unfortunately, we could not solve this integral equation. However, in
principle, there may exist well-behaved solutions which might have
important implications for the moduli space of the $D1/D5$ system.
One possible way is to  find a series solution by iteration which
would be identical to an expansion in powers of $q_1$.  
Secondly, there may be a smooth
solution away from the delta function source. For this purpose, we set
the right hand side of the equation (\ref{rd}) to zero. Then using the
solution for $H_p(r)$ which decays as $r\to\infty$,
(\ref{t}) becomes
\be\label{h2}
H_2=1+q_2\int_0^\infty dp\, c_p\,(yr)^{-1}\,J_1(py)\,K_\nu(pr).  
\ee
At this point, the constant $c_p$ is completely arbitrary (which may
also depend on $q_1$). However, it
should satisfy the following two conditions for a localized
solution. Obviously, (\ref{h2}) should yield a finite $D1$-brane charge
which can be calculated from
\be
\int_{\Sigma} * (dt\wedge dx\wedge dH_2^{-1})
\ee
where $*$ is the Hodge dual and the integral is taken over a
7-dimensional closed surface $\Sigma$ surrounding the $D1$-brane which can be
taken as ($\lim_{y\to\infty} y^3\Omega_3
d^4r+\lim_{r\to\infty}r^3\hat{\Omega}_3d^4 y$), where $\Omega_3$ and
$\hat{\Omega}_3$ are the unit spheres in $\vec{y}$ and $\vec{r}$ spaces,
respectively. The other condition on $c_p$ is that for $q_1=0$,
i.e. $\nu=1$, (\ref{h2}) should give a single $D1$-brane solution.    
However, it turns out to be quite difficult to satisfy both
conditions. For example, it is easy to see that choosing  $c_p=p^3$,
(\ref{h2}) gives $H_2\sim 1+1/(y^2+r^2)^3$ when $q_1=0$ which
is precisely the harmonic function for a single $D1$-brane. Moreover,
$D1$-brane is localized inside the D5-brane i.e. $H_2\to 1$ as
$y\to\infty$. Nevertheless,  the metric has a pathologic divergence as
one approaches the $D5$-brane horizon at $r=0$. To see this let us
consider the integral (\ref{h2}) for large $p$. In this case, $\nu\sim
p \sqrt{q_1}$. For fixed $r$, the modified Bessel function has the
following limiting behavior 
\be\label{limk}
\lim_{\to\infty} K_p(pr)=\sqrt{\frac{\pi}{2p}}\,(1+r^2)^{-1/4}\,e^{-p\eta(r)},
\ee
where $\eta(r)=\sqrt{1+r^2}+\ln r-\ln(1+\sqrt{1+r^2})$. One can see that
there is a positive constant $b$ (which depends on the $D5$-brane
charge $q_1$) such that $\eta>0$ when $r>b$, $\eta<0$ when $r<b$ and
$\eta=0$ when $r=b$. Therefore, the integral (\ref{h2}) converges for
$r>b$ but diverges when $r\leq b$. Note that this is similar to a delta
function type singularity. Due to this pathologic behavior, the total
$D1$-charge diverges.   

\
\

\leftline{${\bf n=1:}$} 

\
\

Eq. (\ref{rd}) can be solved in terms of confluent hypergeometric
functions $U(a,b,r)$ and $M(a,b,r)$. The solution which decays at
large $r$ and regular at $r=0$ can be written as
\be\label{k2}
H_p(r)=\begin{cases}{c_p(r_0)\,e^{-pr}\,U(1+q_1\,p/2\,,2\,,2pr)\hs{6.5}r>r_0,\cr\cr d_p(r_0)\,e^{-pr}\,M(1+q_1\,p/2\,,2\,,2pr)\hs{5}r<r_0.}\end{cases}
\ee
The continuity and discontinuity conditions at $r=r_0$ give
\bea
c_p(r_0)\,U=d_p(r_0)\,M,\\
c_p(r_0)\,(2p)\,\,W\{M,U\}\,=\,q_2\,r_0^{-2}\,e^{pr_0}\,M. 
\eea
where $U$ and $M$ have the same arguments given in (\ref{k2}) and $W$ is
the Wronskian. From the last relation $c_p(r_0)$ can be fixed as
\be
c_p(r_0)=-q_2\,(2p)\,\Gamma(1+q_1p/2)\,M\,e^{-pr_0}.
\ee
Unlike $D1/D5$ case, the constant $c_p$ has a smooth $r_0\to0$ limit
in which it becomes (up to an irrelevant numerical factor)
\be
c_p=q_2\,q_1\,p^2\,\Gamma(q_1\,p/2).
\ee

Now we focus on specific examples. For $D3\perp D5(2)$ it is possible
to delocalize $D3$ or $D5$ branes.   
When $D5$-brane is delocalized inside $D3$-brane, $D5$-brane has the
world-volume coordinates $(\vec{x},\vec{y})$ and $D3$-brane has 
$(\vec{x},\vec{z})$. $H_1$ is the harmonic function of the
$D5$-brane. In this case, $m=3$ and the $\theta$ integral in (\ref{t})
can be calculated easily, which results 
\be\label{h2d3}
H_2=1+q_2\int_0^\infty dp\,p^3\,q_1\,\Gamma(q_1\,p/2)
\,y^{-1}\,\sin(py)\,e^{-pr}\,U(1+q_1\,p/2\,,2\,,2pr).
\ee
Note that, as $y\to\infty$, $H_2\to 1$, which means that 
$D3$-branes are localized inside
$D5$-branes. On the other hand, as $q_1\to \infty$ we have
\bea
H_2&=&1+q_2\int_0^\infty\,dp\, p^2 \,y^{-1}\,\sin(py)\,e^{-pr}\,U(1,2,2pr),\nonumber\\
&=&1+\frac{2q_2}{(r^2+y^2)^2},\label{ai}
\eea
which is precisely the single $D3$-brane solution. To obtain the near
horizon geometry, we use the fact 
\be\label{lim}
\lim_{a\to\infty}\Gamma(1+a-b)\,U(a, b , z/a)
=2 \,z^{\frac{1}{2}-\frac{1}{2}b}\,K_{b-1}(2\sqrt{z}), 
\ee 
and the three dimensional Fourier transform of $K_1$. Defining a new
radial coordinate $\rho^2=q_1 r$ and sending $q_1\to\infty$ while keeping
$\rho$ fixed (which is the near horizon limit) we obtain
\be\label{nh}
H_2=\,q_1\,\frac{6\pi\,q_2}{(y^2+4\rho^2)^{5/2}}.
\ee 
The overall $q_1$ factor can be scaled away in the metric (this is
standard in taking near horizon limits) and this is exactly the
near horizon solution constructed in \cite{youm} and \cite{fay3}. Therefore,
(\ref{h2d3}) gives a background smoothly interpolating  between the 
asymptotically flat and near horizon regions. To see this more
explicitly, one can numerically integrate (\ref{h2d3}). Let us define 
\be\label{sf}
f(r)=k\,r^{5/2}\,\left[H_2(y=0,r)-1\right],
\ee 
where $k$ is a normalization constant. From figure 2, it is possible
to see the behavior of the function $H_2(y=0,r)$ both in the near
horizon and asymptotic infinity which is clearly consistent with 
(\ref{nh}) and (\ref{ai}). 

\begin{figure}
\centerline{
\includegraphics[width=8.0cm]{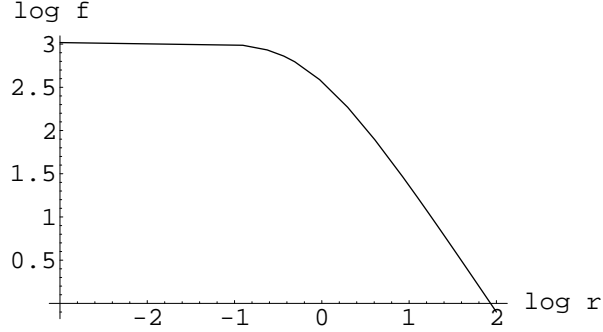}}
\caption{Log-Log plot of the function $f(r)$ (\ref{sf}) for $D3\perp D5(2)$
intersection when $D5$-brane is delocalized along $D3$.}
\end{figure}

In the $D3\perp D5(2)$ intersection when $D3$-brane is delocalized
instead of $D5$-brane, $H_1$ in (\ref{h1}) becomes 
the harmonic function of the $D3$-brane
which has the world-volume coordinates $(\vec{x},\vec{y})$. It is easy
to see that the space transverse to $D5$-brane located inside the $D3$-brane is
one-dimensional thus we have $m=1$. From the first line of (\ref{t})
one obtains
\be\label{h2d4}
H_2=1+q_2\int_0^\infty dp\,p^2\,q_1\,\Gamma(q_1\,p/2)
\,\cos(py)\,e^{-pr}\,U(1+q_1\,p/2\,,2\,,2pr).
\ee
In this solution, delocalization of $D5$-branes inside $D3$-branes,
i.e. the fact that $y\to\infty$, $H_2\to 1$, is guaranteed by the
Riemann-Lebesgue theorem.     
On the other hand, it is easy to see that as $q_1\to 0$
one obtains $H_2=1+q_2/(y^2+r^2)$ which gives the solution
for a single $D5$-brane. To obtain the near horizon limit, we define
$\rho^2=q_1 r$, let $q_1\to0$ while keeping $\rho$ fixed and use (\ref{lim})
to get 
\be
H_2=\frac{2\pi\,q_2}{(y^2+4\rho^2)^{3/2}}.
\ee 
In this expression 
an overall factor of $q_1$ is ignored. Thus (\ref{h2d4}) gives a
solution which interpolates between the asymptotically flat and near
horizon regions. 

Finally, we consider $M5\perp M5 (3)$ intersection in
$D=11$. (The same results also apply to $D4\perp D4(2)$
intersection of type IIA theory). Let us remind that one of the
harmonic functions is given by (\ref{h1}) with $n=1$ corresponding to
a smeared $M5$-brane. The relative transverse space of the other
$M5$-brane located inside the smeared one is two-dimensional. Thus $m=2$
and $H_2$ can be calculated from (\ref{t}) to give
\be\label{h2m5}
H_2=1+q_2\int_0^\infty dp\,p^3\,q_1\,\Gamma(q_1\,p/2)
\,J_0(py)\,e^{-pr}\,U(1+q_1\,p/2\,,2\,,2pr).
\ee
As $y\to\infty$, $H_2\to 1$ hence one of the $M5$-branes is
localized inside the other one. On the other hand, it is easy to see
that as $q_1\to 0$ we have $H_2=1+q_2/2(r^2+y^2)^3$ which is the
solution for a single $M5$-brane. Taking the near
horizon limit by keeping $\rho^2=q_1 r$ fixed as $q_1\to \infty$ 
we obtain (ignoring an overall $q_1$ factor)
\be
H_2=\frac{8\,q_2}{(y^2+4\rho^2)^{2}}.
\ee 
This shows that the solution given by the integral (\ref{h2m5}) smoothly
interpolates between the asymptotically flat and near horizon regions. 

From these examples we see that when the overall transverse space is
three dimensional (which corresponds to $n=1$) it is possible to
obtain smooth solutions in an integral form for partially localized brane
intersections. Therefore, for higher dimensions with $n>1$, it
is possible to smear some directions in the overall transverse space
and reduce the problem to the $n=1$ case. For instance in $D1/D5$
system smearing one direction we get
\be\label{ssmear}
H_2=1+q_2\int_0^\infty dp\,p^4\,q_1\,\Gamma(q_1\,p/2)
\,y^{-1}\,J_1(py)\,e^{-pr}\,U(1+q_1\,p/2\,,2\,,2pr).
\ee
In the near horizon limit defined by $q_1\to\infty$ with fixed
$\rho^2=q_1 r$, we get 
\be
H_2=\frac{32 q_2}{(4\rho^2+y^2)^3}
\ee
which is in agreement with the previously constructed solution given
in \cite{youm}. 

Another way of reducing the power of $r$ in $H_1$ is to consider other
Ricci flat spaces in the transverse part, however this may not be
sufficient alone. For example, for $D1/D5$, one can replace
four-dimensional flat $\vec{r}$ coordinates in (\ref{met1}) with a
Taub-NUT space. Note that no-go theorem does not apply with this
modification.  In this case, the field equations (\ref{1})-(\ref{3})
become  
\bea
\nabla^2_{TN}H_1&=&q_1\,\delta_{TN},\label{t1}\\
(\nabla^2_{TN}+H_1\partial^2_{\vec{y}})H_2&=&q_2\,\delta(\vec{y})\,\delta_{TN},\label{t2}
\eea 
where $\nabla^2_{TN}$ is the Laplacian and $\delta_{TN}$ is the
covariant delta function of the Taub-NUT space which has
the metric
\be
ds^2=\left[1+\frac{2m}{r}\right](dr^2+r^2(d\theta^2+\sin^2\theta d\phi^2))+
\left[1+\frac{2m}{r}\right]^{-1}(4m)^2(d\psi+\frac{1}{2}\cos\theta d\phi)^2.
\ee
For $H_1=H_1(r)$, away from the source (\ref{t1}) becomes
\be
\left[1+\frac{2m}{r}\right]^{-1}\left(\frac{\partial^2}{\partial
r^2}+\frac{2}{r}\frac{\partial}{\partial r}\right) H_1 = 0. 
\ee
This has the solution 
\be\label{qr}
H_1=1+\frac{q_1}{r},
\ee
which precisely obeys (\ref{t1}) with the source term. Now, recall
that $r$ dependence of $H_1$ was $1/r^2$ when the transverse space was
flat. So we achieved our goal and reduced the its power by one. To find
the harmonic function $H_2$, we first put $D1$-brane at $r=r_0$
in Taub-NUT space. Writing $H_2$ as in (\ref{t}), (\ref{t2}) becomes
\be\label{trd}
\left[\frac{d^2}{dr^2}+\frac{2}{r}\frac{d}{dr}-p^2(1+\frac{q_1}{r})(1+\frac{2m}{r})\right]H_p(r)=\,q_2\,\frac{\delta(r-r_0)}{r^2}.
\ee
This can be solved in terms of confluent hypergeometric functions, and
the solution which decays at large $r$ and regular at $r=0$ can be
found as 
\be H_p(r)=\begin{cases}{c_p(r_0)\,e^{-pr}\,r^{-1+\frac{\mu}{2}}\,\,U(m
p+\frac{q_1p+\mu}{2},\mu\,,2pr),\hs{6.5} r>r_0,\cr\cr
d_p(r_0)\,e^{-pr}\,r^{-1+\frac{\mu}{2}}\,\,M(m
p+\frac{q_1p+\mu}{2},\mu\,,2pr),\hs{5} r<r_0,}\end{cases}
\ee
where $\mu=1+\sqrt{1+8mp^2q_1}$. Using the conditions imposed by the
delta function source, it is easy to obtain
\be
c_p(r_0)=\,q_2\,\frac{\Gamma[mp+\frac{q_1p+\mu}{2}]}{\Gamma[\mu]}\,r_0^{-1+\mu/2}\,p^{\mu-1}\,M\,e^{-pr_0}
\ee
In the $r_0\to0$ limit, we have $c_p\to0$ implying spontaneous
delocalization. So, even though the $r$ dependence of $H_1$ in
(\ref{qr}) is lowered by using Taub-NUT space, still it is not
possible to construct a localized $D1\perp D5(1)$ intersection.  

\section{Conclusions}

In this paper we obtained partially localized supergravity solutions 
for $D3 \perp D5 (2)$, $D4 \perp D4 (2)$ and $M5 \perp M5 (3)$
intersections where the overall transverse space is three
dimensional. It is clear that, as in the case of $D2/D6$ intersection
studied in \cite{hashimoto}, our solutions exhibit richer behavior in
the  decoupling limit compared to the completely delocalized or partially
localized but near-horizon solutions \cite{youm}. 

When $n>2$, we could not succeed in solving the radial differential equation. 
Yet the delocalization phenomenon  is expected to occur
\cite{mar2, mar3}. For these cases smearing the overall transverse
dimensions until $n=1$ is an option.  In principle, intersections with
$n\leq 0$ can also be analyzed as above. However, since the asymptotic
geometry is not flat they are not considered in this paper.

For intersections with four dimensional transverse space, the primary
example being $D1\perp D5(1)$, we observed that the method fails,
implying a delocalization which is consistent with the no-go theorems
\cite{mar1, mar2, mar3}.  To overcome this problem we highlighted
two possible ways. Namely, one can solve the integral equation (\ref{sint})
or find a suitable $c_p$ in (\ref{h2}). However these seem to be quite
difficult to come up with.
On the other hand, smearing one transverse dimension 
we obtained a valid supergravity solution (\ref{ssmear}). 
The field theoretic meaning of neither this nor the  near horizon
version given in \cite{youm} is not clear to us. 
This needs  further investigation. We also tried to construct a
localized solution by replacing  the flat transverse space with
Taub-NUT which unfortunately did not improve the situation. It would
be interesting to consider other Ricci flat manifolds.  

Recently $D3 \perp D5 (2)$ intersection has received a lot of interest after 
\cite{karch}. In the approach that we employed we were forced to
delocalize one of the branes. Although this may still be useful for
the purposes of \cite{karch}, a fully localized solution would
probably be more appropriate.

Finally, in \cite{hashimoto}, $D2/D6$ intersection was obtained by
starting from an $M2$-brane which contained Taub-NUT space in the
transverse part. Similarly, $D4/D6$ system can be studied by
considering an $M5$-brane whose two of the world-volume coordinates
embedded holomorphically into a Taub-NUT space \cite{has2, cher}. It
would be interesting to construct this solution which might give some
clue for a more general intersection ansatz.

\begin{acknowledgments}

We would like to thank A. Hashimoto for an e-mail correspondence on 
$D1/D5$ intersection. We also would like to thank T. Rador for
discussions.

\end{acknowledgments}

\end{document}